\documentclass[aip,reprint,a4paper]{revtex4-1}
\usepackage[utf8]{inputenc}
\usepackage[colorlinks=true,linkcolor=blue,citecolor=blue,urlcolor=blue]{hyperref}
\usepackage{amsfonts}
\usepackage{amsmath}
\usepackage{graphicx}
\usepackage{multirow}

\usepackage{epigraph}

\begin{document}


\title{Interpretations of quantum theory: A map of madness}


\author{Ad\'an Cabello}
\affiliation{Departamento de F\'{\i}sica Aplicada II, Universidad de Sevilla, E-41012 Sevilla, Spain.}


\epigraph{\textit{``Knowing the problem is 90\% of solving it''.}}{Bill Evans.}



\bigskip

\bigskip


\begin{abstract}
Motivated by some recent news \cite{Hensen15}, a journalist asks a group of physicists: ``What's the meaning of the violation of Bell's inequality?'' One physicist answers: ``It means that non-locality is an established fact''. Another says: ``There is no non-locality; the message is that measurement outcomes are irreducibly random''. A third one says: ``It cannot be answered simply on purely physical grounds, the answer requires an act of metaphysical judgement''\footnote{All of our characters are fictitious. However, it might be interesting to compare their points of view with those in Refs.~\cite{Gisin12,Zeilinger05,Polkinghorne14}, respectively}. Puzzled by the answers, the journalist keeps asking questions about quantum theory: ``What is teleported in quantum teleportation?'' ``How does a quantum computer really work?'' Shockingly, for each of these questions, the journalist obtains a variety of answers which, in many cases, are mutually exclusive. At the end of the day, the journalist asks: ``How do you plan to make progress if, after 90 years of quantum theory, you still don't know what it {\em means}? How can you possibly identify the {\em physical} principles of quantum theory or expand quantum theory into gravity if you don't agree on what quantum theory {\em is about}?'' Here we argue that it is becoming urgent to solve this too long lasting problem. For that, we point out that the interpretations of quantum theory are, essentially, of two types and that these two types are so radically different that there {\em must be} experiments that, when analyzed outside the framework of quantum theory, lead to different empirically testable predictions. Arguably, even if these experiments do not end the discussion, they will add new elements to the list of strange properties that some interpretations must have, therefore they will indirectly support those interpretations that do not need to have all these strange properties.
\end{abstract}

\maketitle




\begin{table*}[!ht]
\begin{center}
\footnotesize
\begin{tabular}{|l|l|l|}
  \cline{2-3} \multicolumn{1}{r|}{} & \multicolumn{1}{c|}{$\psi$-Ontic} & \multicolumn{1}{c|}{$\psi$-Epistemic} \\
  \hline \multirow{5}{*}{\begin{tabular}{c}Type-I\\ (intrinsic realism)\end{tabular}}
  & Bohmian mechanics \cite{Bohm52,Goldstein13} & Einstein \cite{Einstein36} \\
  & Many worlds \cite{Everett57,Vaidman15} & Ballentine \cite{Ballentine70} \\
  & Modal \cite{vanFraassen72,LD12} & Consistent histories \cite{Griffiths84,Griffiths14} \\
  & Bell's ``beables'' \cite{Bell76} & Spekkens \cite{Spekkens07}\\
  & Collapse theories$^*$ \cite{GRW86,Ghirardi11} & \\
  \hline
\end{tabular}
\begin{tabular}{|l|l|l|}
  \cline{2-3} \multicolumn{1}{r|}{} & \multicolumn{1}{c|}{About knowledge} & \multicolumn{1}{c|}{About belief} \\
  \hline {\multirow{6}{*}{\begin{tabular}{c}Type-II\\ (participatory realism)\end{tabular}}}
  & Copenhagen \cite{Bohr98,Faye14} & QBism \cite{Fuchs10,FS13,FMS14}\\
  & Wheeler \cite{Wheeler83,Wheeler94} & \\
  & Relational \cite{Kochen85,Rovelli96} & \\
  & Zeilinger \cite{Zeilinger99,Zeilinger05} & \\
  & No ``interpretation'' \cite{FP00} & \\
  & Brukner \cite{Brukner15} & \\
  \hline
\end{tabular}
\end{center}
\caption{\label{Table1}{\bf Some interpretations of quantum theory classified according to whether they view probabilities of measurement outcomes as determined or not by intrinsic properties of the observed system.} Type-I interpretations are defined as those in which the probabilities of measurement outcomes are determined by intrinsic properties of the observed system. Type-I interpretations can be ``$\psi$-ontic'' \cite{HS10}, if they view the quantum state as an intrinsic property of the observed system, or ``$\psi$-epistemic'' \cite{HS10}, if they view the quantum state as representing knowledge of an underlying objective reality in a sense somewhat analogous to that in which a state in classical statistical mechanics assigns a probability distribution to points in phase space. ``Type-II interpretations'' are defined as those which do not view the probabilities of measurement outcomes of quantum theory as determined by intrinsic properties of the observed system. Type-II interpretations do not deny the existence of an objective world but, according to them, quantum theory does not deal directly with intrinsic properties of the observed system, but with the experiences an observer or agent has of the observed system. Type-II interpretations can be ``about knowledge'' if they view the quantum state as an observer's knowledge about the results of future experiments, or ``about belief'' if they view the quantum state as an agent's expectations about the results of future actions. This table is based on a similar table made by Leifer \cite{Leifer14}, but has many significant differences and incorporates suggestions from many colleagues (see acknowledgements). The term ``participatory realism'' (inspired by Wheeler's ``participatory universe'' \cite{Wheeler94}) was suggested by Fuchs \cite{Fuchs16}. $*$: Collapse theories modify or supplement the unitary formalism of quantum theory; therefore, they are not pure interpretations.}
\end{table*}


As Mermin points out, ``quantum theory is the most useful and powerful theory physicists have ever devised. Yet today, nearly 90 years after its formulation, disagreement about the meaning of the theory is stronger than ever. New interpretations appear every day. None ever disappear'' \cite{Mermin12}. This situation is odd and is arguably an obstacle for scientific progress, or at least for a certain kind of scientific progress. The periodic efforts of listing and comparing the increasing number of interpretations \cite{Belinfante73,Jammer74,Bub97,Dickson98} show that there is something persistent since the formulation of quantum theory: Interpretations are essentially of two types; those that view quantum probabilities of measurement outcomes as determined by intrinsic properties of the world and those that do not. Here we call them ``type-I'' and ``type-II'', respectively. In Table~\ref{Table1} some interpretations are classified according to this criterion and some extra details are given.

Two observations can be made in the light of Table~\ref{Table1}.
\begin{description}
\item[Observation one] The extended belief that both types of interpretations ``yield the same empirical consequences'' and therefore ``the choice between them cannot be made simply on purely physical grounds but it requires an act of metaphysical judgement'' \cite{Polkinghorne14} is arguably wrong. The two types are so different that is very unlikely that distinguishing between them is forever out of reach of scientific method. In fact, by making some reasonable assumptions it can be shown \cite{CGGLW15} that type-I interpretations must have more strange properties than those suggested by previous approaches \cite{KS67,Bell64,PBR12}.

\item[Observation two] The proposed principles for reconstructing quantum theory \cite{Hardy01,Hardy11,DB11,MM11,CDP11,Cabello13,BMU14,CY14} are neutral with respect to interpretations. This may be a drawback. Although these approaches can lead to the mathematical structure of the theory, the need to remain neutral may be an obstacle for identifying {\em physical} principles. This becomes evident when the proposed principles are examined in the light of some specific interpretations (e.g., QBism \cite{Fuchs10,FS13,FMS14}). Then, not all of them are equally important: Some simply follow from the assumed interpretational framework; only a few of them give some physical insight. This suggests that there may be a bonus in non-neutral reconstructions of quantum theory. This will be developed elsewhere.
\end{description}
This Table and these observations should be taken as motivations for further work. The fact that they may be controversial by themselves, and the interest showed by many colleagues, justify presenting them separately from any of these works.


{\bf Acknowledgements:} This work was supported by the FQXi large grant project ``The Nature of Information in Sequential Quantum Measurements'' and project FIS2014-60843-P (MINECO, Spain) with FEDER funds. I thank D.\ Z.\ Albert, M.\ Ara\'ujo, L.~E.~Ballentine, H.\ R.\ Brown, \v{C}.\ Brukner, J.\ Bub, G.\ Chiribella, D.\ Dieks, C.\ A.\ Fuchs, R.\ B.\ Griffiths, M.\ Kleinmann, J.-{\AA}.\ Larsson, M.\ Leifer, O.\ Lombardi, N.~D.~Mermin, M.\ P.\ M\"uller, R.\ Schack, C.\ Timpson, L.~Vaidman, D.\ Wallace, A.\ G.\ White, and A.\ Zeilinger for conversations and suggestions for improving the table.\\


\end{document}